\documentclass{ws-procs9x6}

\begin{document}

\title{The strong coupling constant at low $Q^2$}

\author{
\vspace{-0.8cm}
A. Deur}

\address{
Thomas Jefferson National Accelerator Facility\\
12000 Jefferson Avenue, Newport News, VA 23606, USA\\ 
E-mail: deurpam@jlab.org}

\maketitle

\abstracts{
\vspace{-0.3cm}
We extract an effective strong coupling constant using low-$Q^2$ data and 
sum rules. Its behavior is established over the full $Q^2$-range and is 
compared to calculations based on lattice QCD,
Schwinger-Dyson equations and a quark model. Although the connection between 
all these quantities is not known yet, the results are surprisingly 
alike. Such a similitude may be related to quark-hadron duality.}
\vspace{-1.3cm}
\section{The strong coupling constant}
A peculiar feature of strong interaction is asymptotic freedom: quark-quark 
interactions grow weaker with decreasing distances. Asymptotic freedom is 
expressed in the vanishing of the QCD coupling constant,$\alpha_s(Q^2)$, at 
large $Q^2$. Conversely, the fact that $\alpha_s(Q^2)$, as calculated in pQCD, 
becomes large when $Q^2 \rightarrow \Lambda_{QCD}^{2}$ is often linked to 
quark confinement. Since it is not expected that pQCD holds at the 
confinement scale and since the condition $\alpha_s(Q^2) \rightarrow 
\infty$ when $Q \rightarrow \Lambda_{QCD}$ is far from necessary to assure 
confinement\cite{Dok}, it is interesting to study $\alpha_s(Q^2)$ in the large 
distance domain.

Experimentally, moments of structure functions are convenient objects to 
extract $\alpha_s$. Among them, $\Gamma_{1}^{p-n}$ is the simplest to use. In 
pQCD, it is linked to the axial charge of the nucleon, $g_{A}$, by the Bjorken 
sum rule:
\vspace{-.15cm}
\begin{eqnarray} 
\Gamma_{1}^{p-n}\equiv\int_{0}^{1}dx(g_{1}^{p}(x)-g_{1}^{n}(x))=\frac{1}{6}
g_{A} [ 1-\frac{\alpha_{s}}{\pi}-3.58
\left(\frac{\alpha_{s}}{\pi}\right)^{2}\\
-20.21\left(\frac{\alpha_{s}}{\pi}\right)^{3}
-130.0\left(\frac{\alpha_{s}} {\pi}\right)^{4}-893.38\left(\frac{\alpha_{s}}
{\pi}\right)^{5}]+ \sum_{i=2}^\infty {\mu^{p-n}_{2i} \over Q^{2i-2}},
\nonumber
\end{eqnarray}
\label{eq:bj}
\noindent
where $g_{1}^{p}$($g_1^{n}$) is the first spin structure function for 
the proton(neutron). The $\mu_t(Q^2)/Q^{t-2}$ are higher twist 
corrections and become important at lower $Q^2$. This series, usually 
truncated to leading twist and to 3rd order, can be used to fit 
experimental data and 
to extract $\alpha_s$. The higher twists can be computed with 
non-perturbative models or can be extracted from data, 
although with limited precision at the moment\cite{momrev}. This imprecise 
knowledge and the break down of pQCD at low $Q^2$ prevent 
\emph{a priori} the extraction of $\alpha_s$ at low $Q^2$. However, an 
\emph{effective} strong coupling constants was defined by 
Grunberg\cite{Grunberg} in which higher twists and higher order QCD 
radiative corrections are 
incorporated. Eq.~\ref{eq:bj} becomes by definition:
\vspace{-.15cm}
\begin{eqnarray}
\Gamma_{1}^{p-n}\equiv \frac{1}{6}g_{A} [ 1-\frac{\alpha_{s,g_1}}{\pi} ]. 
\label{eqn:bjltlo}
\end{eqnarray}
\noindent
This definition yields many advantages: the coupling constant is extractable 
at any $Q^2$, is 
well-behaved when  $Q^2 \rightarrow \Lambda_{QCD}$, is not renormalization 
scheme (RS) dependent and is analytic when crossing quark thresholds. The price
to pay for such benefits is that it becomes process-dependent (hence the 
subscript $g_1$ in Eq.~\ref{eqn:bjltlo}). However, as pointed out by Brodsky 
\emph{et al.}\cite{Brodsky}, effective couplings can be related to each 
other, at least in the pQCD domain, by ``commensurate scale equations''. 
These relate, using different $Q^2$ scales, observables without RS or scale 
ambiguity. Thus, one effective coupling constant is enough to characterize the 
strong interaction.

Among the possible observables available to define an effective coupling 
constant, $\Gamma_{1}^{p-n}$ has unique advantages. The generalized 
Gerasimov-Drell-Hearn (GDH)\cite{GDH,Ji-GDH} and 
Bjorken sum rules predict $\Gamma_{1}^{p-n}$  at low and large $Q^2$, and 
$\Gamma_{1}^{p-n}$ is experimentally known between these two domains. Hence, 
$\alpha_{s,g_1}$ can be extracted at any $Q^2$. In particular, it has a well 
defined value at $Q^2$=0. 
Furthermore, we will see that $\alpha_{s,g_1}$ might best be suited to be 
compared to the predictions of theories and models.
\vspace{-.5cm}
\section{Experimental determination of $\alpha_{s,g_1}$}
A measurement of $\Gamma_{1}^{p-n}$ at intermediate $Q^2$ was reported 
recently\cite{JLab} and was used to extract $\alpha_{s,g_1}$\cite{alpha_s_g1}. 
The results are shown by the triangles in Fig.~1, together with 
$\alpha_{s,g_1}$ extracted from SLAC data\cite{SLAC} at $Q^2$=5 GeV$^2$ (open 
square). Note that the elastic contribution is not included in 
$\Gamma_{1}^{p-n}$.
 
$\Gamma_{1}^{p-n}$ is related to the generalized GDH sums:
\begin{eqnarray}
\vspace{-.2cm}
\Gamma_1^{p-n}= \frac{Q^2}{16 \pi^2 \alpha}(GDH^p-GDH^n)
\label{eqn:gdhcons}
\end{eqnarray}
\noindent  
where $\alpha$ is the QED coupling constant. Hence, at $Q^2$=0, 
$\Gamma_1^{p-n}=0$ and 
\begin{eqnarray}
\vspace{-.2cm} 
\alpha_{s,g_1}= \pi. 
\label{eqn:alpha=pi}
\end{eqnarray}
\noindent   
At $Q^2=0$, the GDH sum rule implies: 
\begin{eqnarray}
\vspace{-.2cm}
\Gamma_1^{p-n}= \frac{Q^2}{16 \pi^2 \alpha}(GDH^p-GDH^n)=\frac{-Q^2}{8}
(\frac{\kappa ^2_p}{M_p^2} - \frac{\kappa ^2_n}{M_n^2})
\label{eqn:gdhcons2}
\end{eqnarray}
\noindent  
where $\kappa _p$ ($\kappa _n$) is the proton (neutron) anomalous magnetic 
moment. Combining Eq.~\ref{eqn:bjltlo} and \ref{eqn:gdhcons2}, we get the 
derivative of $\alpha_{s,g_1}$ at $Q^2$=0: 
\begin{eqnarray}
\vspace{-.2cm}
\frac{d \alpha_{s,g_1}} {dQ^2}= \frac{3 \pi}{4g_A} \times 
(\frac{\kappa ^2_n}{M_n^2} - \frac{\kappa ^2_p}{M_p^2}).
\label{eqn:gdhcons3}
\end{eqnarray}
\noindent    
Relations \ref{eqn:alpha=pi} and \ref{eqn:gdhcons3} constrain 
$\alpha_{s,g_1}$ at low $Q^2$ (dashed line
in Fig.~1). At large $Q^2$, $\Gamma_1^{p-n}$ can be estimated using 
Eq.~\ref{eq:bj} at leading twist and $\alpha_{s}$ calculated with pQCD.
$\alpha_{s,g_1}$ can be subsequently extracted (gray band).

These data and sum rules give $\alpha_{s,g_1}(Q^2)$ at any $Q^2$. A similar 
result is obtained using a model of 
$\Gamma_1^{p-n}$ and Eq.~\ref{eqn:bjltlo} (dotted line). The 
Burkert-Ioffe\cite{BI} model is used because of its good match 
with data. 

One can compare our result to effective coupling constants extracted using
different processes. $\alpha_{s,\tau}$ was extracted from $\tau$-decay 
data\cite{Brodsky2} from the OPAL experiment (inverted triangle). It is
compatible with $\alpha_{s,g_1}$. The Gross-Llewellyn Smith sum 
rule\cite{GLS} (GLS) can be used to form $\alpha_{s,F_3}$. The
sum rule relates the number of valence quarks in the hadron, $n_v$, to the 
structure function $F_{3}(Q^{2},x)$. At leading twist, it reads:
\begin{eqnarray}
\vspace{-.2cm}
\int_{0}^{1}F_{3}(Q^{2},x)dx=n_{v}\left[1-\frac{\alpha_{\rm{s}}(Q^{2})}
{\pi}-3.58\left(\frac{\alpha_{\rm{s}}(Q^{2})}{\pi}\right)^{2}
-20.21\left(\frac{\alpha_{\rm{s}}(Q^{2})}{\pi}\right)^{3}\right].
\end{eqnarray}
\label{eqn:GLS}
\noindent
We expect $\alpha_{s, \rm{F_3}}=\alpha_{s, \rm{g_1}}$ at high $Q^2$, since 
the $Q^2$-dependence of 
Eq.~\ref{eq:bj} and~7 at leading twist are identical. The GLS sum 
was measured by the CCFR collaboration\cite{CCFR} and the resulting 
$\alpha_{s,F_3}$ is shown by the star symbols.

\begin{figure}[ht!]
\begin{center}
\centerline{\includegraphics[scale=0.4, angle=0]{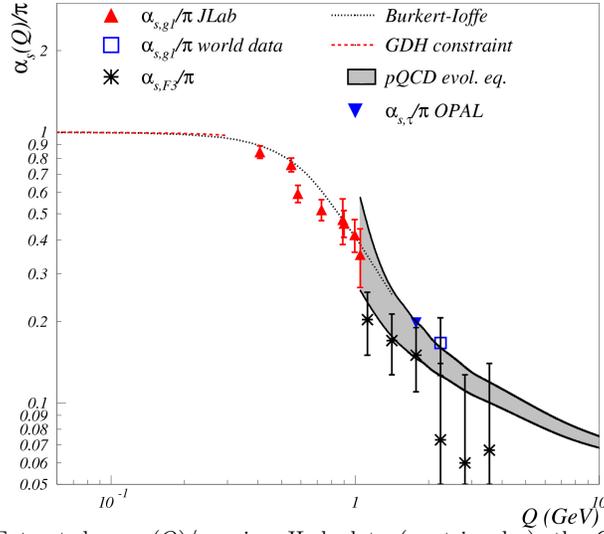}}
\end{center}
\vspace{-1cm}
\caption{
Extracted $\alpha_{s,g_1}(Q)/\pi$ using JLab data (up triangles), 
the GLS sum rule (stars), the world 
$\Gamma^{p-n}_1$ data (open square), the Bjorken sum rule (gray band) and 
the Burkert-Ioffe Model. $\alpha_{s, \tau}(Q)/\pi$ from OPAL is given by the 
reversed triangle. The dashed line is the GDH constrain on the derivative of 
$\alpha_{s,g_1}/ \pi$ at $Q^2$=0.
\vspace{-0.5cm}
}
\label{fig:alpha}
\end{figure}
\vspace{-.5cm}
\section{Comparison with theory}
Just like effective coupling constants extracted 
experimentally, there are also many possible theory definitions for the 
coupling constant and, contrarily to the experimental quantities, the relations
between the various definitions are not well known. Furthermore, the 
connection between the experimental and the theoretical quantities is not 
clear. Hence, the remainder of this paper is to be understood as a candid 
comparison of quantities \emph{a priori} defined differently, in order to see 
if they share common features.

Calculations of $\alpha_s$ using Schwinger-Dyson equations (SDE), lattice QCD 
or quark models are available. Different SDE results
are shown in Fig.~2. The pioneering result of Cornwall\cite{Cornwall} 
is shown by the blue band in the top left panel. The more
recent SDE results from Fisher \emph{et al.}, Bloch \emph{et al.}, Maris and
Tandy, and Bhagwat \emph{et al.} are shown in top left, top right, bottom left 
and bottom left panels respectively. There is a good match between the data 
and the result from Fisher~\emph{et al.} and a fair match with the curve from 
Bloch~\emph{et al.} The results 
from Maris-Tandy, Bhagwat~\emph{et al.} and Cornwall do not match the 
data. The Godfrey and Isgur curve in the top right panel of Fig.~2 
is the coupling constant used in the framework of hadron 
spectroscopy\cite{Godfrey_Isgur}. $Q^2$-behavior of coupling constants can also
be compared regardless of their absolute magnitudes by normalizing 
them to $\pi$ at $Q^2=0$ (These curves are not shown here). The Godfrey-Isgur, 
Cornwall and Fisher~\emph{et al.} $Q^2$-behavior match well the data. The 
normalized curves from Maris-Tandy, Bloch~\emph{et al.} and 
Bhagwat~\emph{et al.} are slightly below the data (by typically one sigma) 
for $Q > 0.6$ GeV. 

Gluon bremsstrahlung and vertex corrections contribute to the
running of $\alpha_{s}$. Modern SDE calculations include those\cite{Tandy2} 
but it is \emph{a priori} not the case for the $\alpha_s$ used in the one 
gluon exchange term of the Godfrey and Isgur quark model, or for older 
SDE works. If so, pQCD corrections should be added to these
calculations. The effect of those corrections (\emph{on $\alpha_{s,g_1}$}) is 
given by the ratio of 
$\alpha_{s,g_1}$ extracted using Eq.~2 to $\alpha_{s,g_1}$ extracted using 
Eq.~1 at leading twist. For both Eq.~1 and~2, $\Gamma^{p-n}_1$ is given by a 
model\cite{BI}. Since model and data agree well, no strong model dependence is 
introduced. The difference between results using Eq.~1 up to 4$^{th}$ and 
5$^{th}$ order is taken as the uncertainty 
due to the truncation of the pQCD series. The resulting $\alpha_s$ are shown 
in the bottom right panel of Fig.~2.

Finally, we can compare lattice QCD data to our results. Many lattice results
are available and are in general consistent. We chose to
compare with the results of Furui and Nakajima\cite{Furui}, see bottom left
panel in Fig.~2. They match well the data. The lowest $Q^2$ point is 
afflicted by finite size effect and should be ignored.

The match between our data and the various 
calculations might be surprising since these quantities are defined 
differently. We can try to understand this fact. Choosing $\Gamma_1^{p-n}$ 
minimizes the r\^ole of resonances, in particular
it fully cancels the $\Delta_{1232}$ contribution which usually dominates
the moments at low $Q^2$. By furthermore excluding the elastic contribution, 
we obtain a quantity for which coherent reactions (elastic and resonances) are 
suppressed and we are back to a DIS-like case in which the 
interpretation is straightforward. One can also possibly invoke the 
phenomenon of quark-hadron duality to explain why the extraction of 
$\alpha_{s,g_1}$, using a formalism developed for DIS\cite{Brodsky2}, seems to 
also work at lower $Q^2$.
\vspace{-0.5cm}
\section{Conclusion}
We have extracted, using JLab data at low $Q^2$ together with sum 
rules, an effective strong coupling constant at any $Q^2$. A 
striking feature is its loss of $Q^2$-dependence
at low $Q^2$. We compared our result to SDE and lattice QCD 
calculations and to a coupling constant used in a quark model. Despite the 
unclear relation between these various coupling constants, data and 
calculations match in most cases, especially for relative 
$Q^2$-dependences. This could be linked to quark-hadron 
duality.
\vspace{-0.5cm}
\section*{Acknowledgments}
This work is supported by the U.S. Department of Energy (DOE). 
The Southeastern Universities Research 
Association  (SURA) operates the Thomas Jefferson National Accelerator
Facility for the DOE under  contract DE-AC05-84ER40150.
\vspace{-0.5cm}

\begin{figure}[ht!]
\begin{center}
\centerline{\includegraphics[scale=0.5, angle=0]{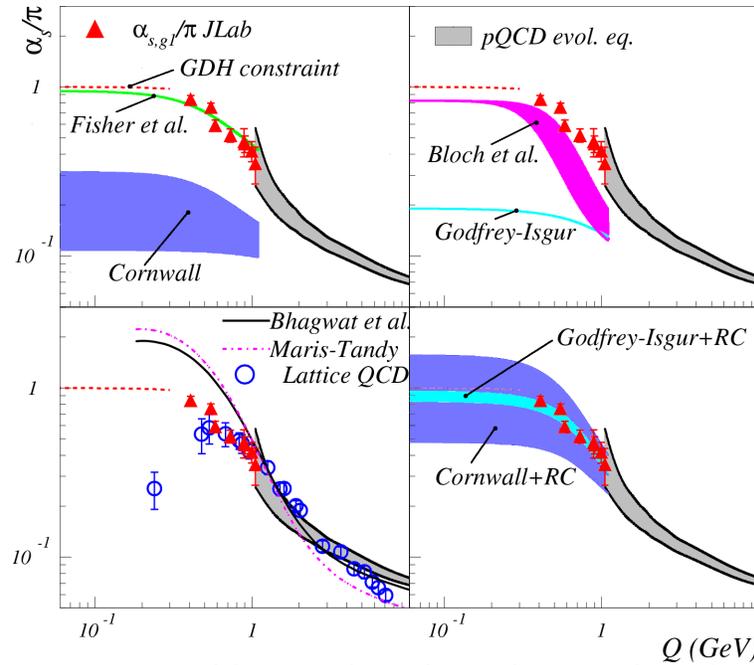}}
\end{center}
\vspace{-1cm}
\caption{$\alpha_{s,g_1}$ extracted from JLab data and sum rules compared to 
various calculations: top left panel: SDE calculations from 
Fisher~\emph{et al.} and Cornwall; top right panel: Bloch~\emph{et al.} (SDE) 
and Godfrey-Isgur (quark model);  bottom left: Furui and Nakajima 
(lattice QCD), Maris-Tandy (SDE) and Bhagwat~\emph{et al.} (SDE); bottom 
right: the Godfrey-Isgur and Cornwall results with pQCD radiative corrections 
added.
}
\label{fig:alpha2}
\vspace{-0.5cm}
\end{figure}

\end{document}